\documentclass[twocolumn,showpacs,prl,superscriptaddress,floatfix]{revtex4}
\usepackage{amsmath,amssymb}
\usepackage{graphicx}

\def\(({\left(}
\def\)){\right)}                       
\def\[[{\left[}
\def\]]{\right]}    

\newcommand{\be}{\begin{equation}}
\newcommand{\ee}{\end{equation}}
\newcommand{\bea}{\begin{eqnarray}}
\newcommand{\eea}{\end{eqnarray}}

\begin{document}

\title{A Lattice Model for Colloidal Gels and Glasses}
  
\author{Florent Krzakala} \affiliation{CNRS; ESPCI, 10 rue Vauquelin,
  UMR 7083 Gulliver, Paris, France 75005, PCT}

\author{Marco Tarzia} \affiliation{Institut de Physique Th\'eorique,
  Orme des Merisiers --- CEA Saclay, 91191 Gif Sur Yvette Cedex,
  France}

\author{Lenka Zdeborov\'a} \affiliation{ Universit\'e Paris-Sud,
  LPTMS, UMR8626, B\^{a}t.~100, Universit\'e Paris-Sud 91405 Orsay
  cedex} \affiliation{CNRS, LPTMS, UMR8626, B\^{a}t.~100, Universit\'e
  Paris-Sud 91405 Orsay cedex}

\begin{abstract}
  We study a lattice model of attractive colloids. It is exactly
  solvable on sparse random graphs. As the pressure and temperature are
  varied it reproduces many characteristic phenomena of liquids,
  glasses and colloidal systems such as ideal gel formation,
  liquid-glass phase coexistence, jamming, or the reentrance of the
  glass transition.
\end{abstract}
  
  
\maketitle

There is a growing experimental and theoretical interest in the study
of colloidal suspensions of mesoscopic particles immersed in a
polymeric solution. Their properties can be controlled precisely via
chemical or physical manipulations, and the inter-particle interaction tuned from a very short-range depletion-induced attraction to a
long-range Coulombic repulsion~\cite{inter}. This makes colloids
important both in the terms of basic scientific research and
industrial applications~\cite{ReviewAll}. Colloidal suspensions also
offer new scenarios of dynamically arrested structures, which are the
main focus of this work.

At high densities these systems exhibit the same glassy dynamics as a
system of hard spheres~\cite{pusey}. The lack of available space at
densities $\rho \approx 0.58$ gives rise to a phase called {\it
  repulsive glass}. If a short-range attraction is present, ergodicity
may also be broken, resulting in the so-called {\it attractive
  glass}. When the range of attraction is less than $\approx 10 \%$ of
the sphere diameter, the interplay between these two mechanisms can
lead to a remarkable reentrance of the glass transition line, with the
(energetically driven) attractive glass dominating at low temperature
and the (entropically driven) repulsive glass at high
temperature. This reentrance was first predicted by the mode coupling
theory and later confirmed by experiments and
simulations~\cite{AGmct}.  Interestingly at low temperatures the
attractive glass transition moves to much lower densities. Such a
low-density arrested state is often identified as a colloidal
gel~\cite{AttractiveGlass=Gel} -- a low-density disordered arrested
state which does not flow but possess solid-like properties such as a
yield stress. However, the dynamical arrest at low densities and its
exact relation to the glass transition is still very poorly
understood, and our work also sheds light on this aspect.

In the low density -- low temperature regime, a system of hard-core
particles with a short-range attraction undergoes a phase
separation between a colloid-rich and a colloid-poor
phase~\cite{Hill}. The interplay between phase separation and gelation
is also a widely discussed topic. Typically, when quenching from high
temperature, the phase separation boundary is crossed first. When
the glass transition is crossed at lower temperatures, the
spinodal decomposition induces the formation of dense arrested
regions, a situation called {\it arrested phase
  separation}~\cite{arrested}. Another phenomenon can be
obtained by reducing the phases-coexisting regions to even lower densities
and temperatures, by decreasing the average coordination number of
particles, as proposed in~\cite{DeKo,Patchy}. Under these conditions the phase separation does not intervene in a wide region of densities and the system can be equilibrated by annealing down to very low temperatures,
leading to a state of matter called {\it ideal gel}~\cite{ideal_gel}.

Theoretical progress has been hampered by the lack of microscopic
models able to roughly locate and describe the colloidal gel
transition. In this letter we discuss an exactly solvable yet
realistic model characterized by valence-limited interactions that
generalizes the lattice glass model introduced by Biroli and M\'ezard
(BM) in~\cite{BiroliMezard}. The model is motivated by the classical
short-range depletion attraction but also by recent studies of a
family of patchy \cite{Patchy} or limited-valency potentials
\cite{Valence}. It allows to reproduce the behavior of a broad variety
of complex materials that are usually simulated at high computer
cost. Under different conditions we observed in particular the
liquid-glass coexistence, the reentrant glass transition and the
formation of an ideal gel.  This leads to an unifying description of
the structural arrest in different amorphous and disordered systems,
ranging from hard-spheres to colloids and gels.

\paragraph*{The model ---}
We consider a system of particles on a lattice. Each site $i$ carries
$n_i=0$ or $n_i=1$ particles, with a short-range attractive
interaction giving a unit energy gain to all pairs of
nearest-neighbors particles.  Motivated by the concepts of geometrical
frustration and valence-limited
interactions~\cite{DeKo,Patchy,Valence}, we follow
BM~\cite{BiroliMezard} and restrict the occupation by a hard
constraint: a particle cannot have more than $\ell$ occupied
neighbors. The partition function (where $\beta$ is the inverse
temperature and $\mu$ the chemical potential) reads
\begin{equation}
  Z(\mu,\beta) = \sum_{{\rm allowed} \, \{n\}} e^{\mu \sum_{i=1}^N n_i + \beta \sum_{<ij>} 
    n_i n_j}, \label{part_sum}
\end{equation}
where the sum is restricted over all configurations $\{n\}$ satisfying
the hard constraints. For $\beta=0$, the attraction 
is turned off and the model reduces to the BM model~\cite{BiroliMezard}.
In real (continuous) space, the presence of the maximum possible
number of neighbors $l$ gives rise to an entropic
loss~\cite{Tarjus}. This effect is, however, inhibited on the lattice
and we thus also consider a variant of the model where each sites with
{\it exactly} $\ell$ occupied neighbors induce a penalty $e^{-s_{p}}$
in eq.~(\ref{part_sum}).  This entropic repulsion is only relevant at
large temperatures and is small compared to the attractive potential
at low temperatures.

\paragraph*{The solution ---}
We shall focus on the exact solution of the model on large $c-$regular
random graphs (where the number of neighborhooding sites,
connectivity, of each vertex is exactly $c>l$). Such graphs are
locally tree-like, i.e., there are no cycles up a distance scaling
like $\log{N}$ from a typical site. The topological disorder and the
geometrical frustration are introduced by very long loops that
disfavor the crystalline order and let emerge more easily the glassy
phases. This corresponds to a mean-field like approach to the problem
\cite{KirkpatrickThirumalai}.

In the low density--high temperature region the Bethe approximation
gives an exact solution on tree-like lattices. For $s_p=0$ (but the
generalization is straightforward) it is obtained from the fixed point
of \bea \psi_{e}=\frac{1}{Z_0}\, , \quad \psi_{s}=
\frac{e^{\mu+\ell\beta}}{Z_0} {c-1\choose \ell} \psi_{u}^{\ell} \psi_{e}^{c-1-l} \, ,\\
\psi_{u}= \frac{e^{\mu}}{Z_0} \sum_{\ell'=0}^{\ell-1} e^{\ell'\beta}
{c-1\choose \ell'} \psi_{u}^{\ell'} \psi_{e}^{c-1-\ell'}\, , \eea
where $\psi_e/\psi_s/\psi_u$ are the probabilities that a node is
empty/saturated/unsaturated when one of its neighborhooding sites is
removed from the graph, $Z_0$ is a normalization constant ensuring
that $\psi_e+\psi_s+\psi_u=1$. The free energy density $J=\log{Z}/N$
then reads
\bea 
J(\mu,\beta) &=&\log{\left[ 1 +  e^{\mu} \sum_{\ell'=0}^{l} e^{\ell'\beta} {c\choose \ell'}
 \psi_{u}^{\ell'} \psi_{e}^{c-\ell'} \right]} \\
\nonumber
&-& \, \frac{c}{2} \log{\left[\psi_{e}^2 + 2 \psi_{e} \((\psi_{u} + \psi_{s}
    \)) + e^{\beta} \psi_{u}^2 \right]}\, . 
\eea
The density of occupied sites is given by $\rho=\partial_\mu
J(\mu,\beta)$.

As the density increases, or the temperature decreases, a dynamical
glass transition arises at $\rho_d(\beta)$ where the free energy landscape divides into
exponentially many valleys~\cite{Cavity}. Each valley corresponds to a
different Gibbs state $\alpha$ with a free energy density
$J_{\alpha}$. This can be recognized from the divergence of the
point-to-set correlation length, or, equivalently, by the emergence of
a nontrivial glass solution described by the one-step replica symmetry
breaking approach~\cite{Cavity,BetheDynamics,Reconstruction,US}. The
complexity $\Sigma(J)$ (sometime called configurational entropy) is
defined as $\Sigma(J)=\log{{\cal N}(J)}/N$, where ${\cal N}(J)$ is the
number of states with free energy density $J$. It can be computed via
its Legendre transform $\Phi(m,\mu,\beta)$ defined by
\begin{equation}
     e^{ N \Phi(m,\mu,\beta) } \equiv \sum_\alpha e^{mNJ_\alpha(\mu,\beta)} = \int {\rm d}J \,  e^{N\big[mJ(\mu,\beta)+\Sigma(J)\big]}\, ,\nonumber
\end{equation}
where the sum runs over all the different Gibbs states~$\alpha$ and
the Legendre parameter $m$ is equivalent to the Parisi parameter in
the replica theory~\cite{Monasson,Cavity}. The cavity method derives a set of self-consistent equations for
$\Phi(m,\mu,\beta)$~\cite{Cavity}. They can be solved analytically in
the limit $\mu \to \infty$~\cite{BiroliMezard}. In general, however,
these equations are solved using the numerical population dynamics
technique of~\cite{Cavity}. The case $m=1$ also allows significant
simplification \cite{Reconstruction}. We also checked
self-consistently, using the method of \cite{US}, that the one-step
replica symmetry breaking solution is stable,
and the presented equilibrium solution is thus generally considered exact.

\paragraph*{The glass and jamming transitions ---}
\begin{figure}
  \hspace{-0.35cm}
  \includegraphics[width=\linewidth]{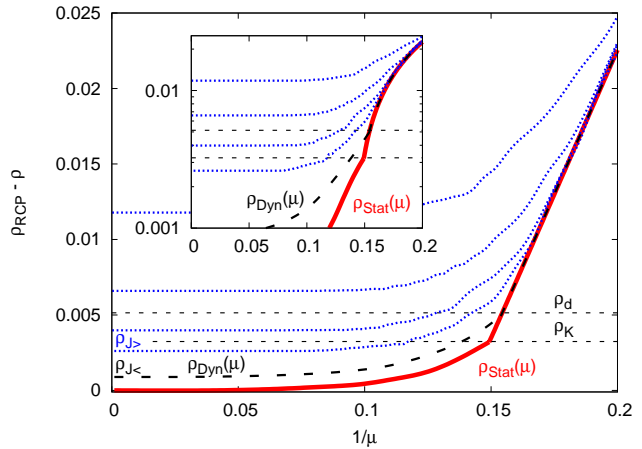}
  \caption{\label{fig1} (color online) High temperature behavior of
    the model with $c=3,\, l=1$ and no entropic penalty. $\rho_{\rm
      RCP} - \rho$ is the difference between the packing fraction, or
    density, and its maximum achievable value (random close packing),
    $\mu$ is the chemical potential. The static equilibrium solution
    (bold/red) undergoes a phase transition at $\rho_K=0.5725$ and
    converges to the densest packing $\rho_{\rm RCP}=0.57574$ at
    $\mu=\infty$. However, the dynamics falls out of equilibrium at
    $\rho_d=0.5708$ and the packing fraction following a slow
    annealing can be increased up to at most $0.5731 \le \rho_J\le
    0.5748$.  The dotted/blue curves are Monte-Carlo annealings for
    $N=10^5$ with rates $d\mu=10^{-4},10^{-5},10^{-6}$ and $10^{-7}$
    (from top to bottom) and the (dashed/black) line $\rho_{\rm Dyn}$
    is the {\it iso-complexity approximation} for the infinitely slow
    annealing (see the text). The inset shows the same plot on a
    logarithmic scale.
    \vspace{-0.5cm}
    }
\end{figure}

\begin{figure*}[!ht]
\begin{center}
\hspace{-0.6cm}
\includegraphics[width=6.4cm]{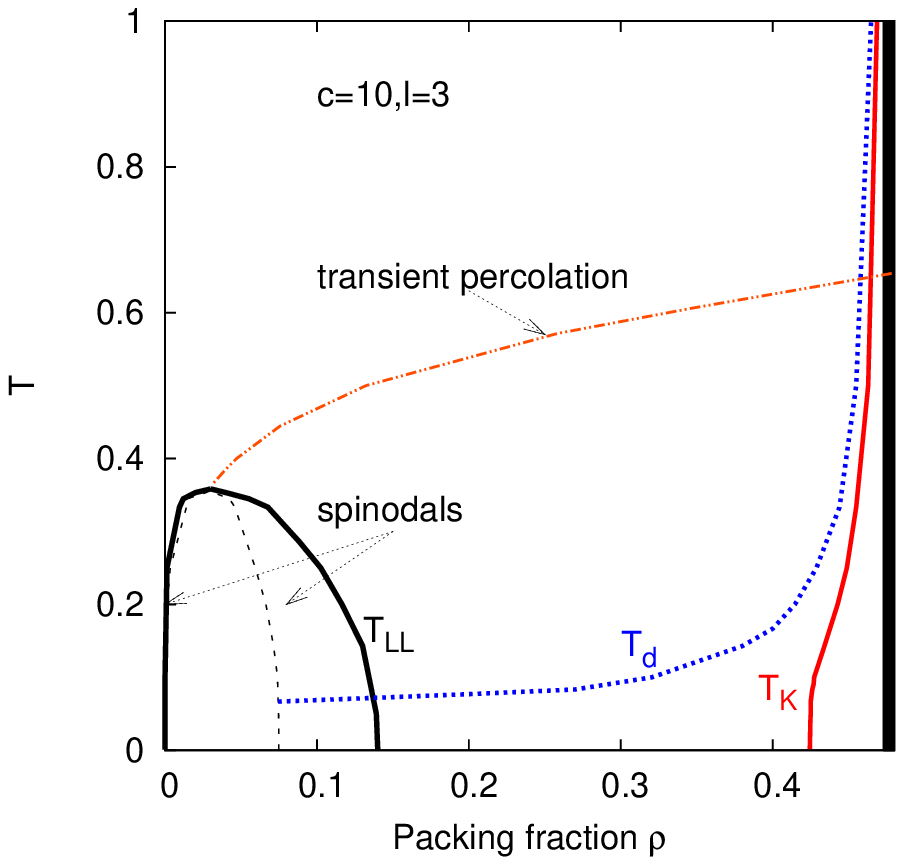}
\hspace{-0.6cm}
\includegraphics[width=6.4cm]{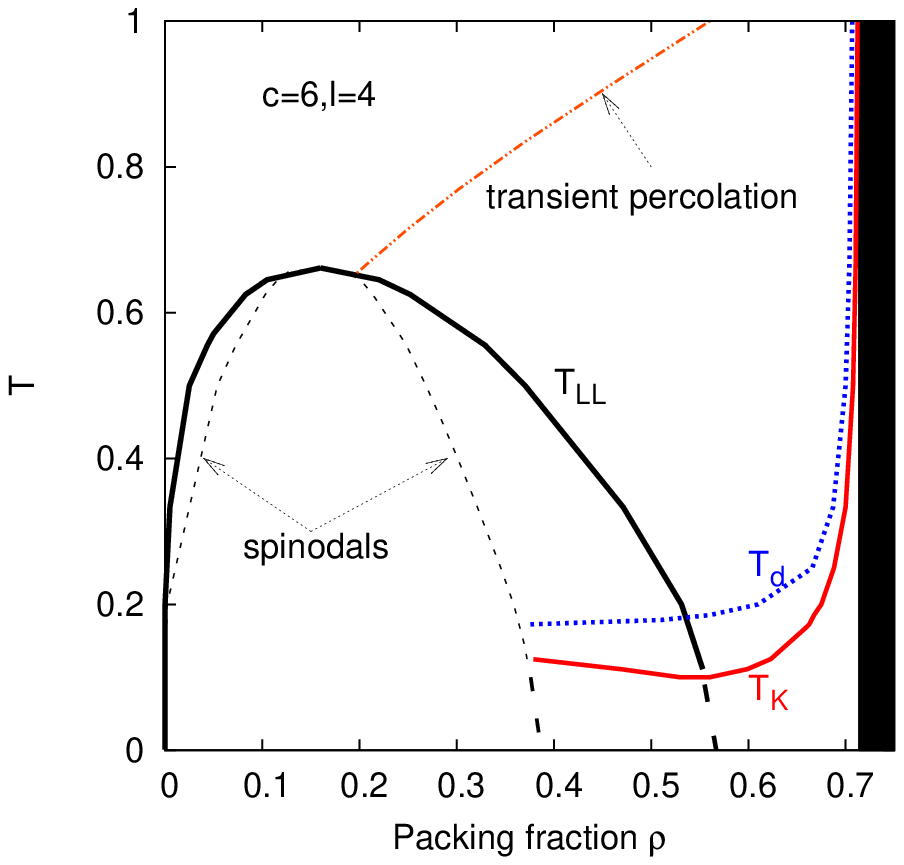}
\hspace{-0.6cm}
\includegraphics[width=6.4cm]{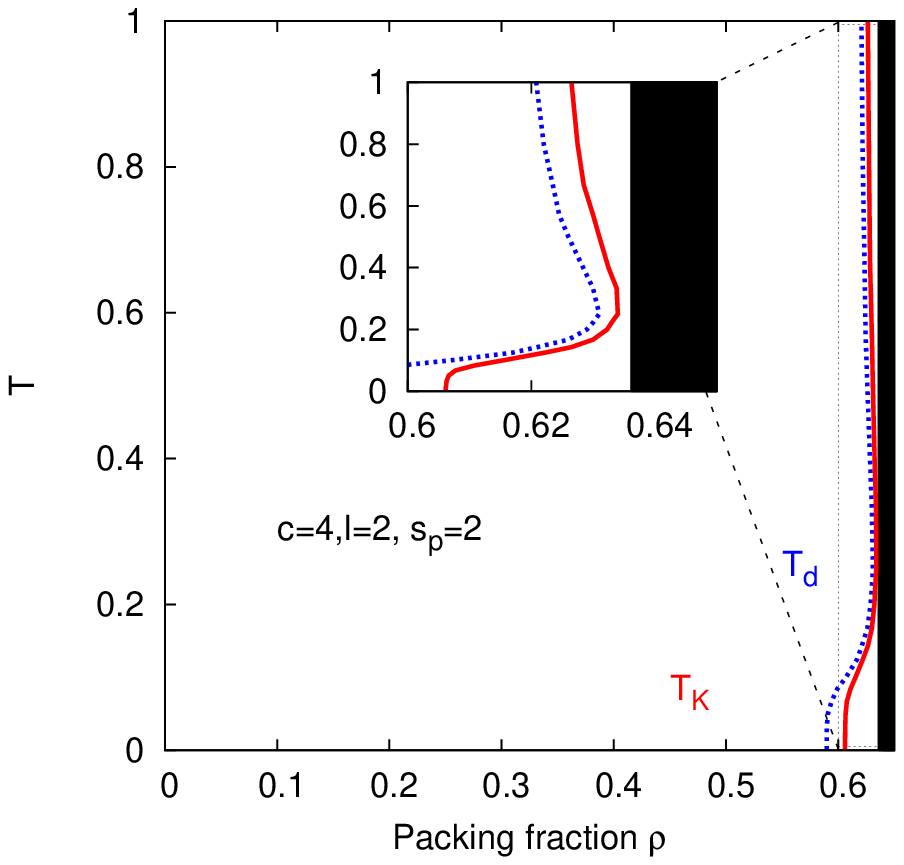}
\end{center}
\caption{(color online) Three representative phase diagrams of the
  model in the temperature ($T$) --- packing fraction ($\rho$) plane.
  The black/continuous curves are the colloid-rich/colloid-poor
  coexistence lines $T_{\rm LL}$, and the corresponding spinodal lines
  are black/dashed curves. The orange/dashed-dotted curves correspond
  to the percolation of physical clusters, and intercept the critical
  point of the phase separation. The dynamical $T_d (\rho)$
  (blue/dotted curve) and the Kauzmann $T_K (\rho)$ (red/continuous
  curve) glass transition lines are also shown. The black filled area
  is beyond the random close packing fraction $\rho_{\rm RCP}$.  {\bf
    Left panel:} $c=10$, $l=3$ and no entropic penalty. An ideal gel
  appears at low enough temperature ($T \lesssim 0.1$) and low volume
  fraction ($0.1 \lesssim \rho \lesssim 0.4$).  {\bf Central panel:}
  $c=6$, $l=4$ and no entropic penalty. A phase coexistence between a
  colloid-poor and an attractive colloid-rich glass phase occurs for
  low temperature ($T \approx 0.05$) resulting in an arrested phase
  separation. {\bf Right panel:} $c=4$, $l=2$ and $s_p=2$.  The glass
  transition lines $T_d (\rho)$ and $T_K (\rho)$ exhibit a reentrant
  shape (see inset), resulting in a glass-liquid-glass transition
  along isochores at densities around $\rho \approx 0.62$.
  \vspace{-0.3cm}
  }
\label{fig2}
\end{figure*}
We first revisit the behavior the of the model in the high temperature
limit ($\beta=0$) where it reduces to the original
BM~\cite{BiroliMezard}, and consider the case $c=3$,$\ell=1$for
concreteness. FIG.~\ref{fig1} depicts in bold/red the difference
between the largest possible density of occupied sites $\rho_{\rm
  RCP}$ and the equilibrium density $\rho$ as a function of the
chemical potential $\mu$. In the equilibrium solution, the model
undergoes a Kauzmann transition at $\rho_K=0.5725$ where the
complexity vanishes and the system becomes a an ideal glass (a
``static'' glass phase)~\cite{KirkpatrickThirumalai}. The density of
the glass can be in principle further increased in the equilibrium
solution, and the densest possible packing is reached at $\rho_{\rm
  RCP}=0.57574$ (Random Close Packing, first computed
in~\cite{BiroliMezard}.)

The dynamical behavior is more involved~\cite{Jorge}. The
equilibration time diverges (as a power law) at the dynamical
transition $\rho_d=0.5708<\rho_K$ where the system is trapped in one
of the exponentially many Gibbs states and falls
out-of-equilibrium. When the pressure is further increased the system
evolves within this state and the density can again be further
increased~\cite{KrzakalaKurchan}, but does not follow the equilibrium
curve. It is not yet known how to describe analytically this process
on random graphs, and in FIG.~\ref{fig1} we thus display the results
of Monte Carlo annealings at different rates. The slowest annealing we
tried reached $\rho \approx 0.5731$. We also computed the {\it
  iso-complexity approximation} curve for an infinitely slow
annealing, where the dynamics is assumed to follow a curve defined by
$\Sigma(\rho_{\rm Dyn} (\mu)) =\Sigma(\rho_d)$~\cite{MontanariRicci}.
This actually yields a lower bound on the infinite pressure density
reached by this procedure.  In our model, the density reached by an
infinitely slow annealing, $\rho_J$, is the analog of the {\it Jamming
  point} \cite{JAM} and we thus obtain $0.5731\ge \rho_J\ge
0.5748$. Note that, as discussed in \cite{KrzakalaKurchan}, $\rho_J$
does not correspond to the densest possible packing $\rho_{\rm RCP}$
neither to the dynamical or Kauzmann density $\rho_d$ or $\rho_K$. In
finite dimensional systems the dynamical transition is smeared out by
activation processes that restore the ergodicity and prevent a
complete freezing of the system~\cite{KirkpatrickThirumalai}. In that
case, an ideal glass transition (if any) should thus arise only at the
Kauzmann point $\rho_K$.

\paragraph*{Phase diagrams ---}
We now discuss the different phase diagrams obtained by varying the
connectivity $c$, the maximum number of occupied neighbors $\ell$, and
the entropic penalty $s_p$. Three representative situations are
depicted in FIG.~\ref{fig2}.

The left panel of FIG.~\ref{fig2} shows a typical phase diagram when
the valency restriction $l$ is small (but larger than $2$) and the
dimension-like parameter $c$ is large. In this situation a
colloid-rich/colloid-poor phase coexistence is located in a narrow
region of small temperatures $T$ and densities $\rho$, in agreement
with~\cite{Patchy}.  The dynamical glass transition line $T_d$ then
bends towards low value of $\rho$ and continues in an almost
horizontal direction. Under these conditions dynamical arrest
phenomena and glassy behavior should emerge at very small volume
fractions. This state of matter is as close as possible to an {\it
  ideal gel}~\cite{ideal_gel,Patchy} which is in our model equivalent
to a glass in the dynamical phase. In contrast, the Kauzmann glass
transition line $T_K$ typically stays almost vertical in the
$T$-$\rho$ plane: a true equilibrium glass transition thus only takes
place at high densities. In a finite dimensional system, where the
dynamical transition is smeared out, the dynamics along isochores
should follow approximatively an Arrhenius dependence, exactly as
found in~\cite{DeKo,Patchy}.  Interestingly enough, the percolation
line is located much above the dynamic arrest (gel) one. This gap is
due to finite life time of bonds at finite temperature, this is common
for colloidal gelation.  Nevertheless, the formation of a spanning
cluster could affect the dynamical behavior of the system, as shown
in~\cite{perco}.

The central panel of FIG.~\ref{fig2} shows a typical phase diagram
when the valency restriction $l$ is large and $c$ small. The
coexistence region is much broader with respect to the previous case.
And both the dynamical and the Kauzmann transition lines bend towards
low densities. An ideal glass state thus emerges for values of $\rho$
much smaller than $\rho_{\rm RCP}$.  The glass transition lines
moreover intercept the spinodal so that a deep quench of the system at
low enough temperatures (e.g., $T \approx 0.05$) into the coexistence
region (e.g., $\rho \approx 0.2$) results in a phase separation
between a colloid-poor and a glassy colloid-rich region.  Under these
conditions the phase separation process is interrupted by the
formation of an (attractive) glass in the dense phase.  This {\it
  arrested phase separation} scenario is one
of the possible routes to colloidal gelation and the formation of a
low density gel state~\cite{arrested}.

The right panel of FIG.~\ref{fig2} shows a case with the valency
restricted to $l=2$. In this case the occupied sites can create only
chains and the phase separation region is absent, this was also
observed in \cite{Patchy}. A more interesting feature shown in this
diagram is the reentrant shape of the dynamical and the Kauzmann
transition lines. The reentrant behavior can be observed when the
short-range attractive potential is in competition with the entropic
repulsion, $s_p$, in agreement with~\cite{AGmct}. At large
temperatures the interaction contributing to the formation of a glass
is the entropic repulsion $s_p$, at low temperature it is the
attractive potential. Note that the reentrant behavior is usually
observed together with the presence of the phase separation region,
and we indeed observe the two effects in the left and central panel if
we set the entropic repulsion $s_p$ to a positive value.  Our example
shows however that the reentrant behavior is not necessarily connected with
the phase separation, nor with a presence of a glass-glass transition,
in agreement with earlier works~\cite{AGmct,NEWREFS}.

\paragraph*{Conclusions ---}
We have studied a lattice model of complex fluids, which can be
solved exactly on random graphs using the cavity method~\cite{Cavity}.
Despite its intrinsic simplicity, it captures many features that arise
in real systems and it illustrates physical properties of arrested
states observed in colloidal suspensions under different
conditions. It reproduces the shape of a variety of phase diagrams,
allowing for a qualitative comparison with experiments.
In the future, it should be interesting to study the coarsening
dynamics associated to the phase separation between the colloid-poor
and the colloid-rich phase and to simulate the dynamics in finite dimension, as
in \cite{BiroliMezard,Coniglio,Lattice}. It is also of interest
to understand if a glass-glass transition can be found for this
class of models.

\acknowledgments We thank A. Coniglio, J. Kurchan,
O. Rivoire, T. Mora, F. Zamponi and E. Zaccarelli for discussions.


\vspace{-0.2cm}

\begin{thebibliography}{99}

\vspace{-0.2cm}

\bibitem{inter} A. Yethiraj and A. V. Blaaderen, Nature {\bf 421}, 513 (2003). 

\bibitem{ReviewAll} V. Trappe {\it et al.}, Nature {\bf 411}, 772
  (2001); F. Sciortino and P. Tartaglia, Advances in Physics {\bf 54},
  471 (2005); W. B. Russel {\it et al.}, {\it Colloidal Dispersions}
  (Cambridge University Press, Cambridge, 1991).

\bibitem{pusey} P. N. Pusey and W. van Megen, Phys. Rev. Lett. {\bf 59}, 2083
(1987).

\bibitem{AGmct} K. A. Dawson {\it et al.}, Phys. Rev. E {\bf 63},
  011401 (2001); K. Pham {\it et al.}, Science {\bf 296} 104 (2002);
  G. Foffi {\it et al.}, Phys. Rev. E {\bf 65}, 050802 (2002). S. Chen
  {\it et al.}, Science {\bf 300}, 619 (2003).

\bibitem{AttractiveGlass=Gel} H. Verduin and J. Dhont,
  J. Coll. Int. Sci. 172, 245 (1995); N. Verhaegh {\it et al.},
  Physica A {\bf 242}, 104 (1997).  S. A. Shah {\it et al.}, J. Phys.:
  Condens. Matter {\bf 15}, 4751 (2003); P. N. Segr\`e {\it et al.},
  Phys. Rev. Lett. {\bf 86}, 6042 (2001).

\bibitem{Hill} Hill T L, {\it An Introduction to Statistical
    thermodynamics}, New York: Dover, 1987.

\bibitem{arrested} P. J. Lu {\it et al}, Nature 453, 499-503 (2008).

\bibitem{DeKo} E. Del Gado and W. Kob, Europhys. Lett. {\bf 72}, 1032
  (2005); {\it ibid}, Phys. Rev. Lett. {\bf 98}, 028303 (2007).

\bibitem{Patchy} E. Zaccarelli {\it et al.}, J. Chem. Phys. {\bf 124},
  124908 (2006); E. Bianchi {\it et al.}, Phys. Rev. Lett. {\bf 97},
  168301 (2006). S. Sastry {\it et al.}, J. Stat. Mech. P12010 (2006).

\bibitem{ideal_gel} F. Sciortino et al.. Computer Physics
  Communications 169 (2005) 166-171.  E. Zaccarelli, J. Phys.:
  Condens. Matter 19, 323101 (2007).

\bibitem{BiroliMezard} G. Biroli and M. M\'ezard,
  Phys. Rev. Lett. {\bf 88}, (2002) 025501. O. Rivoire {\it et al.},
  Eur. Phys.  J. B {\bf 37}, 55 (2004).

\bibitem{Valence} R. J. Speedy and P. G. Debenedetti, Mol. Phys. {\bf 81}, 
237 (1994).


\bibitem{Tarjus} G. Tarjus {\it et al.}, J. Phys.: Condens. Matter {\bf 15}, 
1077 (2003).

\bibitem{KirkpatrickThirumalai} T. Kirkpatrick and D. Thirumalai,
  Phys. Rev.  Lett. {\bf 58}, 2091 (1987); T. Kirkpatrick, P. Wolynes,
  Phys. Rev. A {\bf 35}, 3072 (1987). M. M\'ezard and G. Parisi,
  Phys. Rev. Lett. {\bf 82}, 747 (1999). G. Parisi and F. Zamponi,
  J. Chem. Phys. {\bf 123}, 144501 (2005).

\bibitem{Cavity} M. M\'ezard and G. Parisi, Eur. Phys. J. B {\bf
    20} (2001) 217.

\bibitem{BetheDynamics} A. Montanari and G. Semerjian, 
    J. Stat. Phys. {\bf 124}, 103 (2006). J.-P. Bouchaud and
  G. Biroli, Phys. Rev. B {\bf 72}, 064204 (2005).

\bibitem{Reconstruction} M. M\'ezard and A. Montanari,
  J. Stat. Phys. {\bf 124} 1317 (2006). F. Krzakala {\it et al.},
  Proc. Natl. Acad. Sci.  {\bf 104}, 10318 (2007).

\bibitem{US} F. Krzakala and L. Zdeborov\'a, Europhys. Lett. {\bf 81}
  (2008) 57005. L. Zdeborov\'a and F. Krzakala, Phys. Rev. E {\bf 76} (2007) 031131.

\bibitem{Monasson} R. Monasson, Phys. Rev. Lett. {\bf 75}, 2847 - 2850
  (1995).

\bibitem{Jorge} J~.P. Bouchaud {\it et al.} in {\it Spin Glasses and
    Random Field}, P. Young ed., World Scientific, Singapur (1997).
  L. F. Cugliandolo, in {\it Les Houches Session 77}, J-L Barrat, J
  Dalibard, J Kurchan, M V Feigel'man eds (2002).

\bibitem{KrzakalaKurchan} F. Krzakala and J. Kurchan, Phys. Rev. E {\bf
    76}, 021122 (2007).

\bibitem{MontanariRicci} A. Montanari and F. Ricci-Tersenghi,
  Phys. Rev B {\bf 70}, 134406 (2004).

\bibitem{JAM} B. Lubachevsky and F. Stillinger, J. Stat. Phys. {\bf
    60}, 561 (1990). A. Liu and S. Nagel, Nature {\bf 396}, 21 (1998).

\bibitem{perco} A. Coniglio {\it et al.}, J. Phys.: Condens. Matter {\bf 16},
S4831 (2004); E. Del Gado, A. Fierro, L. de Arcangelis, and A. Coniglio,
Phys. Rev. E {\bf 69}, 051103 (2004).


\bibitem{NEWREFS} G. Foffi {\it et al} Phys. Rev. Lett. 90, 238301
  (2003), P. Kumar {\it et al.} Phys. Rev. E 72, 021501 (2005)).


\bibitem{Coniglio} M. Pica Ciamarra {\it et al.}, Phys. Rev. E {\bf 67},
057105 (2003).

\bibitem{Lattice} E. Marinari and V. Van Kerrebroeck,
  Europhys. Lett. {\bf 73}, 383 (2006), and G. D. McCullagh {\it et
    al.}, Phys. Rev. E {\bf 71}, 030102(R) (2005).


\end{thebibliography}
\end{document}